%% file: eprint.tex
\documentclass[12pt]{article}
\usepackage{graphicx}

\newcommand\pubnumber{CIPANP2018-Arrington}
\newcommand\pubdate{\today}

\def\anl{Physics Division, Argonne National Lab, Lemont, IL, USA}
\def\support{\footnote{This work was supported by the U.S. Department of Energy, Office of Science, Office of Nuclear Physics under contract DE-AC02-06CH11357.}}

\def\Title#1{\begin{center} {\Large #1 } \end{center}}
\def\Author#1{\begin{center}{ \sc #1} \end{center}}
\def\Address#1{\begin{center}{ \it #1} \end{center}}

\newcommand\pubblock{\rightline{\begin{tabular}{l} \pubnumber\\
         \pubdate  \end{tabular}}}
\newenvironment{Abstract}{\begin{quotation}  }{\end{quotation}}
\newenvironment{Presented}{\begin{quotation} \begin{center} 
             PRESENTED AT\end{center}\bigskip 
      \begin{center}\begin{large}}{\end{large}\end{center} \end{quotation}}
\def\Acknowledgements{\bigskip  \bigskip \begin{center} \begin{large}
             \bf ACKNOWLEDGEMENTS \end{large}\end{center}}

\input econfmacros.tex

\begin{document}
\begin{titlepage}
\pubblock

\vfill
\Title{Inclusive Studies of Short-Range Correlations: Overview and New Results}
\vfill
\Author{Zhihong Ye, John Arrington\support}
\Address{\anl}
\vfill
\begin{Abstract}
We present an overview of Short-Range Correlations (SRC) studies using the inclusive measurement of the
electron scattering off nuclei. A brief introduction of the origin of the SRC is given, followed by the survey of
the two-nucleon SRC (2N-SRC) study and its interesting connection to the EMC effect. A discussion of the
three-nucleon SRC study (3N-SRC) measured by the Jefferson Lab's Hall B and Hall C experiments which showed
contradictory results is given and, most importantly, we report a new result from the Hall A E08-014
experiment which was dedicated on studying 3N-SRC. Our high precision $^{4}$He/$^{3}$He cross section ratios at
$x>2$ region do not show a 3N-SRC plateau as predicted by the naive SRC model. To further investigate the 3N-SRC as well as the isospin effect of the SRC, we have designed several approved experiments in Hall A and in Hall C, including the Tritium
experiments using the mirror nuclei ($^{3}$H and $^{3}$He) which are currently running in Hall A.
\end{Abstract}
\vfill
\begin{Presented}
CIPANP 2018\\
Palm Springs, CA, May 29 - June 3, 2018
\end{Presented}
\vfill
\end{titlepage}
\def\thefootnote{\fnsymbol{footnote}}
\setcounter{footnote}{0}

\section{Short-Range Correlations}
\begin{figure}[htb]
\centering
\includegraphics[height=2.4in]{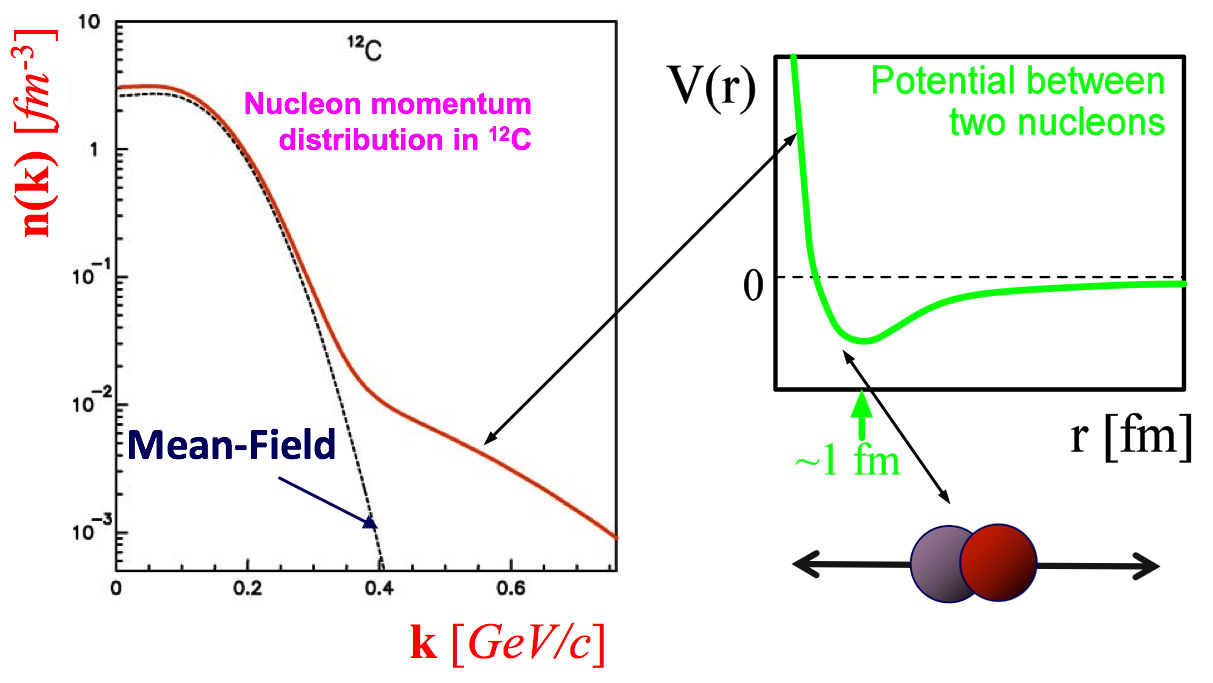}
\caption{Momentum distribution for carbon in the Mean-Field approach and explicitly including the short-range two-body
nucleon interactions. Figure adopted from Ref.~\cite{CiofidegliAtti:1995qe}}
\label{fig:nnforce}
\end{figure}

\begin{figure}[htb]
\centering
\includegraphics[height=2.5in]{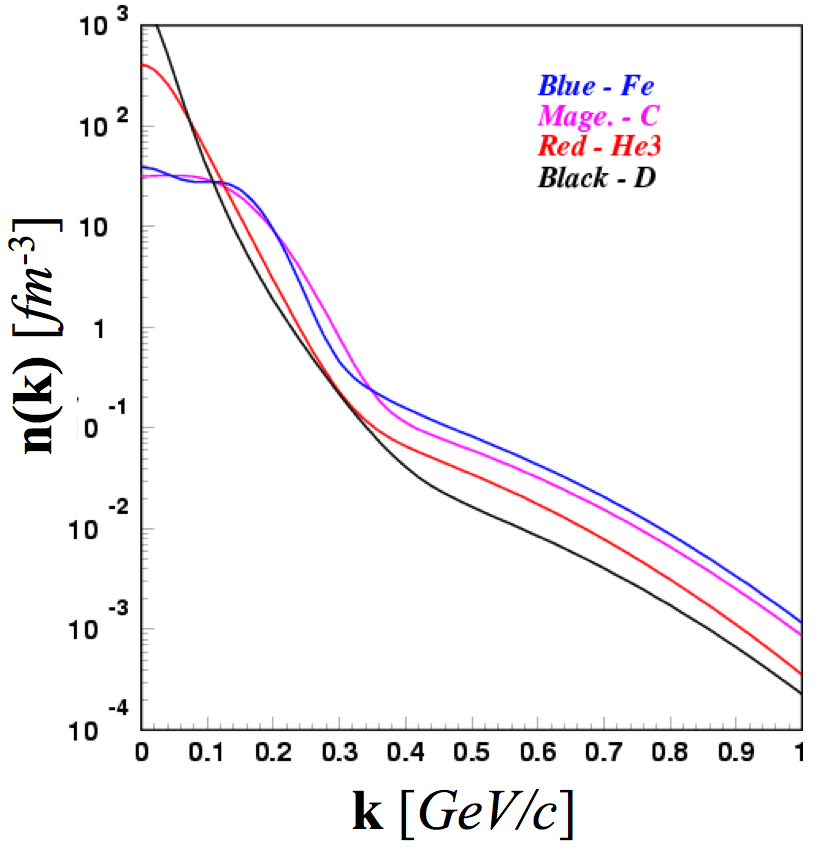}
\caption{Momentum distributions among light to heavy nuclei.}
\label{fig:mom_dis}
\end{figure}

Nuclei are composed of nucleons, i.e. protons and neutrons, whose interactions are relatively weak at typical
inter-nucleon distances ($>$2 fm), but much stronger at short distances ($\sim$1 fm). At small separation, the
nucleons' two-body potential, dominated by tensor force, becomes much stronger until their hard cores start to push
nucleons away from each other, resulting in strong repulsive force, shown in Fig.~\ref{fig:nnforce}. Mean-Field
theory, such as the Shell-Model, has been a successful theory to describe how nucleons move inside a nucleus below
the typical Fermi-momentum ($k_F \approx 250$ MeV/c). There is, however, still significant contribution from
nucleons that carry momentum above the Fermi-motion, indicated by the high momentum tail of a nucleon's momentum
distribution
in a nucleus~\cite{CiofidegliAtti:1995qe}. Even in $^2$H, nearly half of its kinematic energy comes from the 5\% of nucleons
with momentum above the Fermi-momentum. These fast-moving nucleons come mainly from the strong short-range interactions
associated with the tensor force and hard repulsive core of the nucleon-nucleon interaction. As such, these high-momentum
nucleons correspond to highly localized configurations, referred to as Short-Range Correlations (SRC). While
the mean field parts of the momentum distributions reveal strong $A$-dependence, the consequence of SRC is the
A-independent high momentum tails at $k>k_{F}$ in different nuclei, as shown in Fig.~\ref{fig:mom_dis}.

One can perform $A(e,e'p)$ experiments and, in a plane-wave interpretation, extract the momentum distribution of
protons in the nucleus. The contribution of SRCs is isolated by looking for the high-momentum nucleons in the
inititial state generated by the short-range configurations, which appear at high missing momentum. However, such
measurements also include contributions from final-state interactions (FSIs), meson-exchange currents (MEC), and
isobar contributions (IC). Even for measurements at large $Q^2$ values, the FSI contribution tend to be significant 
for measurements at large missing momentum~\cite{Sargsian:2002wc, Arrington:2011xs}. This makes it extremely
difficult to isolate the contribution from SRCs. Despite the limitations of A(e,e'p) measurements, they have been
successfully used to identify scattering events which appear to come from high-momentum protons in the initial
state, allowing a direct test of the SRC picture by looking for the partner nucleon from the SRC with momenta
opposite of the struck nucleon~\cite{Subedi:2008zz, Korover:2014dma}. These data support the SRC model, and also
demonstrated that np pairs dominate the high-momentum component of the nuclear momentum distribution.

\begin{figure}[htb]
\centering
\includegraphics[height=2.3in]{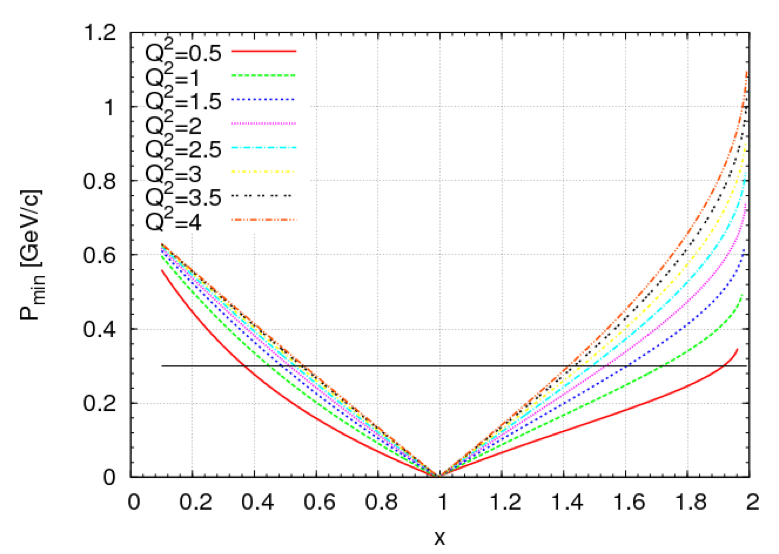}
\caption{Correlation of $Q^2$ and $x$ corresponding to different nucleon initial momenta. Figure from Ref.~\cite{Fomin:2017ydn}}
\label{fig:qe_kin}
\end{figure}

In inclusive scattering, these final state interactions are significantly suppressed. However, one cannot directly reconstruct
the initial momentum of the struck nucleon. Despite this, one can still isolate SRC contributions by measuring the inclusive
cross sections in select kinematic regions, e.g. large $Q^2$ and $x>1.4-1.5$. For these kinematics, FSI and MEC effects
and inelastic contributions are largely suppressed, and at large $x$, the scattering is sensitive only to
scattering from nucleons above a minimum momentum which can be selected to isolate SRCs by choosing appropriate
$Q^2$ and $x$ values, as shown in Fig.~\ref{fig:qe_kin}. This allows us to isolate SRC contributions and compare
their strength and structure in different nuclei. Thus, we can examine the relative contribution of SRCs in different
nuclei and verify their universal character, but we do not have enough information to reconstruct the initial
momentum of the knock-out protons, nor to directly observe the back-to-back motion of the nucleon pairs in
the center of mass frame.

The inclusive QE cross section of electron scattering off a nucleus ($A$) at $x>1$ can be decomposed into contribution
from single-particle states, two-body, three-body, and N-body contributions, in the simple SRC
model~\cite{ Frankfurt:1993sp}, neglecting center-of-mass motion of the N-body configurations, as
\begin{equation}
   \sigma^A_{QE}(x) = \sigma^A_{MF}(x) +a_2 \sigma^A_2(x)+a_3 \sigma^A_3(x) + \cdots ~ ,
   \label{qe_eq}
\end{equation}
where $\sigma^A_{MF}(x)$ denotes the contribution from the mean-field part, while $\sigma^A_2(x)$ ($\sigma^A_3(x)$)
represents the contribution from the 2N-SRC (3N-SRC). $a_2^A$ ($a_3^A$) gives the probability of forming the 2N-SRC
(3N-SRC) configuration and should show the scaling behavior in the range of $1<x<2$ ($2<x<3$).
\begin{figure}[htb]
  \centering
  \includegraphics[height=2.5in]{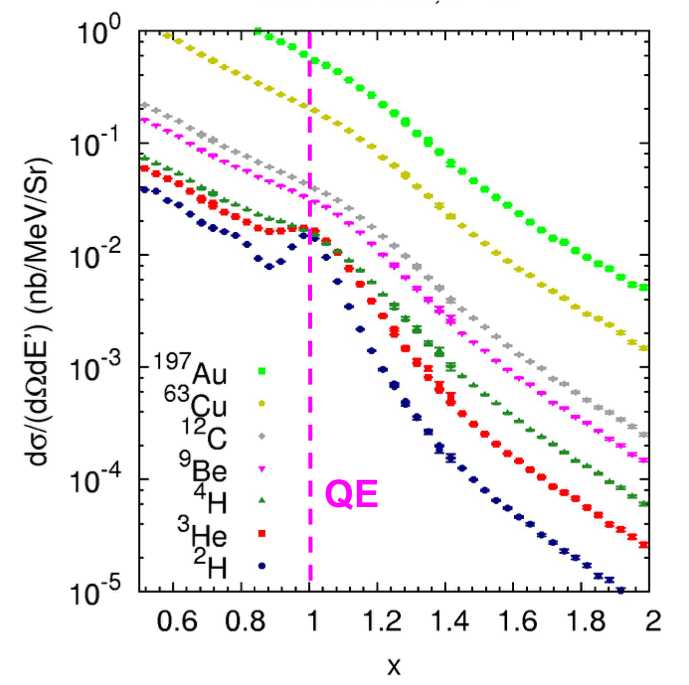}
  \caption{Inclusive quasi-elastic cross section distributions among light to heavy nuclei~\cite{nadia_thesis}.}
  \label{fig:qe_xs}
\end{figure}

If kinematics are selected such that scattering requires initial momenta above the Fermi momentum, then the
mean-field contribution is negligible and the cross section reduces to
\begin{equation}
  \sigma^A_{QE}(x) = a_2 \sigma^A_2(x)+a_3 \sigma^A_3(x) + \cdots ~ .
  \label{qe_eq2}
\end{equation}
This is typically achieved by requiring $Q^2$>1-2~GeV$^2$ and $x>1.4-1.5$. With 3N-SRC contributions expected
to be small below $x$=2, this implies a universal behavior for the cross section in these kinematics, as demonstrated
in the QE cross sections of these nuclei, shown in Fig.~\ref{fig:qe_xs}. Near the QE peak, the cross section has
a strongly $A$-dependent shape, with a narrow, pronounced QE peak for light nuclei, and a washed-out peak for heavier
nuclei. However, above $x\approx$1.4, the cross section shows a universal behavior, with only the normalization
varying with $A$.

\begin{figure}[htb]
  \centering
  \includegraphics[height=2.5in]{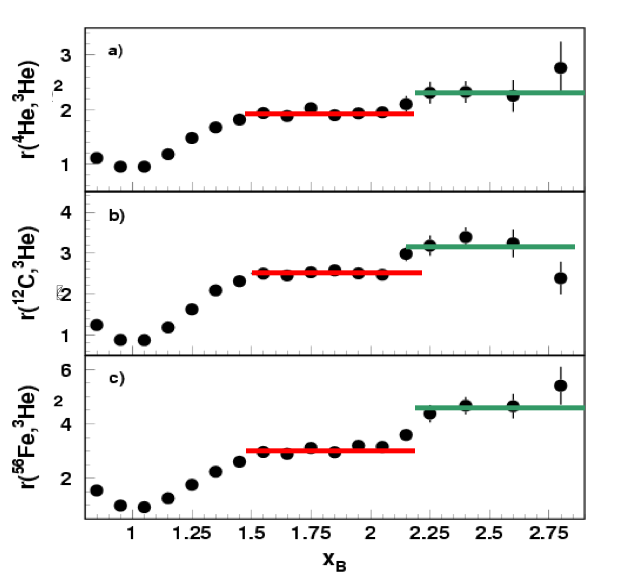}
  \includegraphics[height=2.5in]{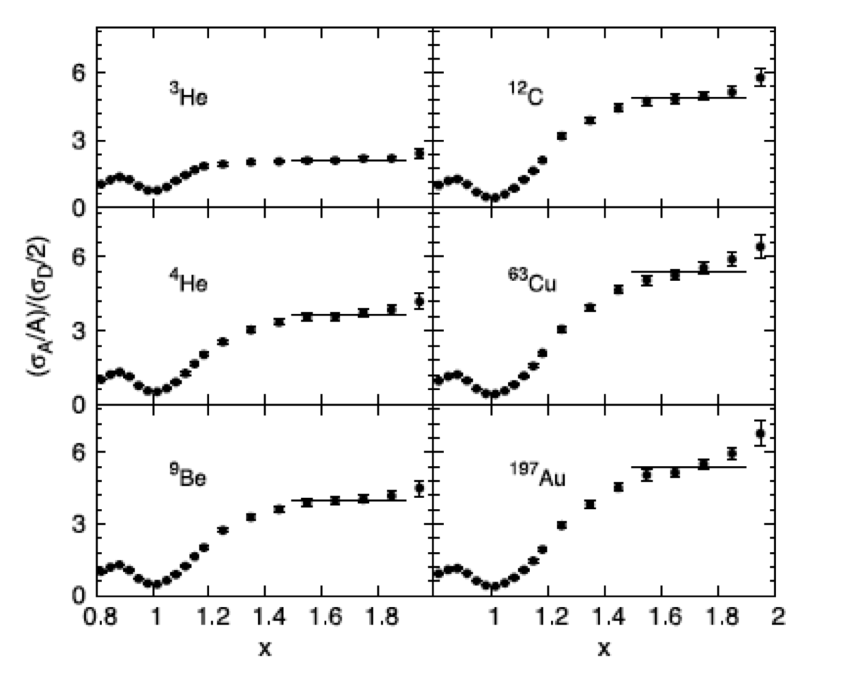}
  \caption{Inclusive cross section ratio of nuclei to $^{3}$He from Hall B CLAS result (left)~\cite{Egiyan:2005hs}
    and nuclei to $^{2}$H from Hall C results (right)~\cite{Fomin:2011ng} }.
  \label{fig:src_hallbc}
\end{figure}

This universal behavior is more clearly seen in Fig.~\ref{fig:src_hallbc}, where ratios of heavy nuclei to
$^2$H and $^3$He are shown from measurements in Halls B and C. The large-$x$ plateau demonstrates the universal
behavior of the cross section, and the ratio corresponds to the relative contribution of 2N-SRCs from the heavy
nucleus relative to $^2$H or $^3$He. Extending this to the region above $x=$2, Eq.~\ref{qe_eq} predicts a
similar plateau for $x>$2, allowing the extraction of $a_2^A$ and $a_3^A$:
\begin{equation}
  a^A_2 = \frac{2}{A}\frac{\sigma^A_2(x)}{\sigma^{^{2}H}_2(x)}, ~~~ a^A_3 = \frac{3}{A}\frac{\sigma^A_3(x)}{\sigma^{^{3}He}_3(x)}.
\label{scaling}
\end{equation}

The contribution of 2N-SRCs has been well studied and shows good agreement between the data from SLAC~\cite{Frankfurt:1993sp}
and from Jefferson Lab's Hall B~\cite{Egiyan:2005hs} and Hall C~\cite{Fomin:2011ng}. Fig.~\ref{fig:src_hallbc} shows a clear
2N-SRC plateau in both the A/$^{3}$He and A/$^{2}$H ratios for $1.5<x<2$, despite the different $Q^2$ ranges and
nuclei in the denominator. The  scaling plateau in the cross section ratio conforms the similarity of the momentum
distributions among nuclei when $k>k_F$ suggested in Fig.~\ref{fig:mom_dis}, while the plateau values
corresponding to $a^A_2$ gives the probability of a nucleon in a 2N-SRC pair. The value of $a_2^A$ changes rapidly with $A$
for light nuclei, but grows slowly for above $A=12$, with heavier nuclei yielding $a_2 \approx 5$. For many years after 
the initial observation of SRCs~\cite{Frankfurt:1993sp}, it was assumed that the contribution of SRCs, measured through
the value of $a_2^A$, scaled with the nuclear density~\cite{Egiyan:2005hs}. This was natural as the typical nucleons
separation, and thus the likelyhood to interact via the short-range parts of the NN potential, increase with the nuclear
density, and was supported by measurements of $a_2^A$ until the data from Ref.~\cite{Fomin:2011ng}. This data demonstrated
showed that $a_2^A$ for $^9$Be deviated significantly from the assumed scaling with the average nuclear density. This was
the first observation that details of nuclear structure had a significant impact on the relative contribution from SRCs.

As noted earlier, the JLab E01-015 $^{12}$C(e,e'pN) experiment demonstrated that np pairs dominate
SRCs~\cite{Subedi:2008zz}, as a consequence of the dominance of the tensor force in at these momenta. While inclusive
measurements combine contributions from protons and neutrons, we can still use inclusive scattering to 
study the isospin structure of SRCs by comparing targets with different isospin structure. The JLab Hall A E08-014
experiment~\cite{e08-014} studied Calcium isotope, $^{48}$Ca and $^{40}$Ca, to test the np-dominance picture
in inclusive scattering, where FSI contributions are expected to be much smaller. An ongoing experiment in Hall A,
E12-11-112~\cite{e12-11-112}, is measuring the inclusive QE cross section ratio of $^{3}$He to $^{3}$H which is
predicted to have much greater sensitivity then measurements using the Calcium isotopes.

\section{2N-SRC and EMC Effect}
\begin{figure}[htb]
\centering
\includegraphics[height=1.5in]{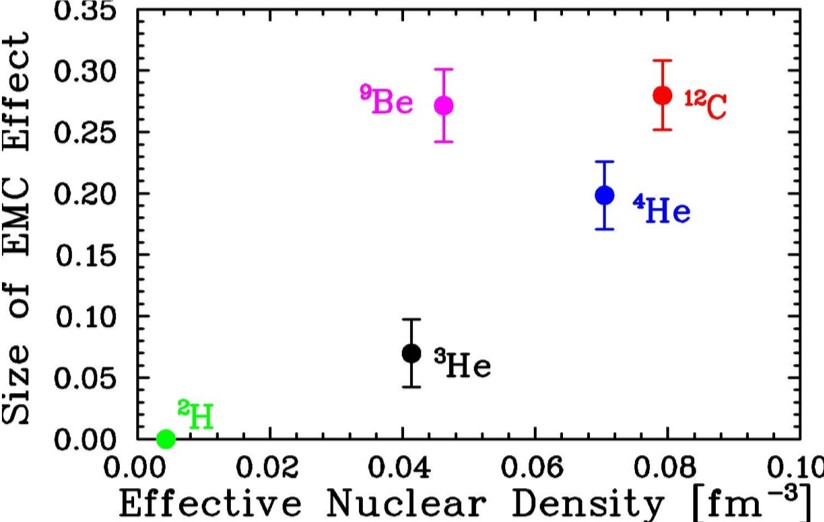}
\includegraphics[height=1.5in]{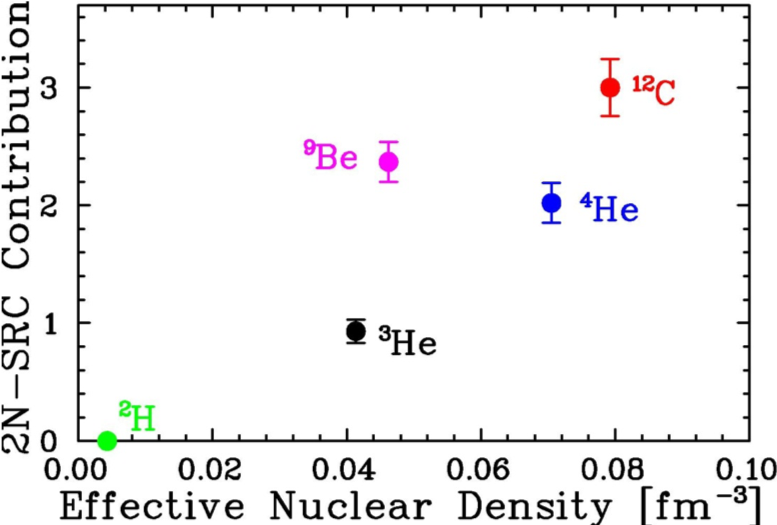}
\caption{Size of the EMC effect~\cite{Seely:2009gt} (left) and contribution from SRCs~\cite{Fomin:2011ng} (right) as a function of nuclear density in 
light nuclei. Figure adopted from Ref.~\cite{Seely:2009gt}.}
\label{fig:emc_src_density}
\end{figure}

The EMC effect was discovered in the 1980s by the European Muon Collaboration at CERN. When using heavy nuclei 
as effective nucleon target to perform high precision measurement of quark parton distribution functions
(PDF), they observed an A-dependent slope in the range of $0.3<x<0.7$ when taking DIS cross section ratio between
nucleus A to deuterium~\cite{Aubert:1983xm}. As with the contribution of SRCs, the EMC effect was observed to scale
with the average density of the nucleus. The most recent measurement of the EMC effect in Hall C at Jefferson Lab
~\cite{Seely:2009gt} made a more precise measurements of the EMC effect in light nuclei up to $^{12}$C. The data 
showed that the size of the EMC effect does not depend simply on the average nuclear density. The $^{9}$Be EMC effect,
as quantified by the slope of the $A$/$^2$H ratios, deviates from the trend observed in other nuclei, as shown in
Fig.~\ref{fig:emc_src_density}. $^{9}$Be is known to have a configuration of two alpha plus a neutron so the observation
suggests that the EMC effect also depends on the local density. This same behavior was later observed in the measurement
of $a_2^A$, also shown in Fig.~\ref{fig:emc_src_density}.

\begin{figure}[htb]
\centering
\includegraphics[height=2.5in]{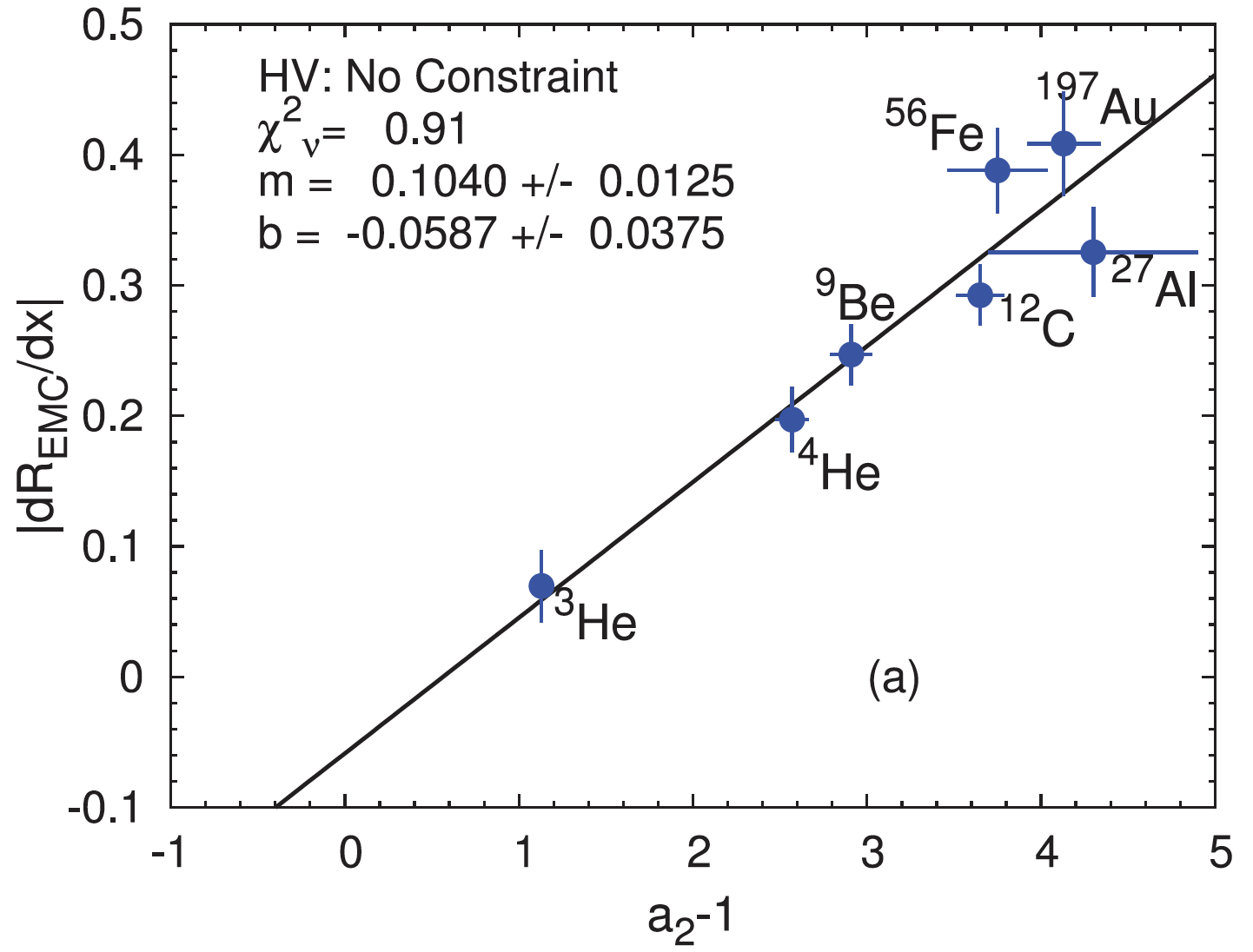}
\caption{The linear correlation between the SRC and the EMC effect. Figure is from~\cite{Arrington:2012ax}.}
\label{fig:emc_src_linear}
\end{figure}

For light nuclei, these can be examined as a function of 
density as calculated from \textit{ab initio} calculations~\cite{wirginga:2013ala}. One can include heavy nuclei in the comparison and avoid
any model dependence by directly comparing the EMC effect and contribution from SRCs, which shows a clear correlation 
between these effects, even for $^9$Be where both effects deviation from the simple density dependence, as shown in
Fig.~\ref{fig:emc_src_linear}. This has led to the suggestion that both the EMC effect and SRCs depend on the same aspect
of nuclear structure, or that one of these is responsible for the other~\cite{Arrington:2012ax}. For example, in the former
case, the short-distance configurations can be taken as the cause for both the SRCs, which arise from the short-distance
interactions, and the EMC effect if it is connected to the local high-density configuration associated with two overlapping
nucleons. For the latter case, it is sometimes assumed that the EMC effect comes from off-shell effects associated with 
high-momentum nucleons, and thus that high-momtenta associated with SRCs is the direct cause of the EMC effect. Existing
data do not provide enough information to separate these (or other) explanations~\cite{Arrington:2012ax}, although such
pictures do make different predictions about other observables, e.g. the flavor-dependence of the EMC
effect~\cite{Arrington:2015wja}.

\section{3N-SRC and New Results}
\begin{figure}[htb]
\centering
\includegraphics[height=2.3in]{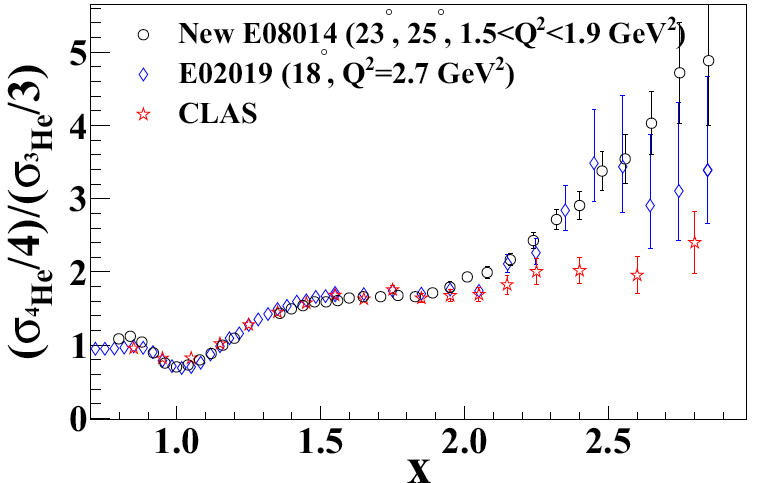}
\caption{New $^{4}$He/$^{3}$He cross section ratio result from the Hall A E08-014 experiment ~\cite{Ye:2017mvo} }.
\label{fig:3nsrc}
\end{figure}

Unlike the 2N-SRC contributions, 3N-SRC contributions have not been clearly identified, and the CLAS data in Hall B
and the E02-019 data in Hall C results didn't show agreement in the $x>2$ region~\cite{Fomin:2011ng}. As shown in
Fig.~\ref{fig:src_hallbc}, the CLAS result show clear plateaus for A/$^{3}$He cross section ratios for $x>2.25$ for
three different nuclei. The E02-019 result, taken at larger $Q^2$ values (roughly 2.8 GeV$^2$, as opposed to $\sim$1.6
GeV$^2$ for the CLAS data) have larger uncertainties but show significantly larger ratios in the $^4$He/$^3$He ratios
for $x>2.3$. This suggests an inconsistency between the data sets or else an unexpectedly large $Q^2$ dependence in
the ratios. One possible explanation of the discrepancy was presented in Ref.~\cite{Higinbotham:2014xna}, suggesting
that due to its limited momentum resolution, the CLAS result has large bin migration effects at large $x$ where the
cross section drops rapidly for $^3$He.

Experiment E08-014~\cite{e08-014} in Hall A performed a dedicated measurement of the $x$ and $Q^2$
dependence of the ratio at $x>2$ with the high-resolution spectrometers for $Q^2$ values close to the CLAS data. The
result~\cite{Ye:2017mvo} is consistent in the 2N-SRC region with the CLAS and E02-019 results, as shown in
Fig.~\ref{fig:3nsrc}. However, the much higher precision result at $x>2$ shows no indication of the 3N-SRC plateau
seen in the CLAS data. This rules out the idea of a large $Q^2$ dependence in the ratio, and is consistent with the
explanation presented in~\cite{Higinbotham:2014xna}. These results show that there is no plateau in the $^4$He/$^3$He
in the 3N-SRC region at these $Q^2$, but does not rule out the presence of significant 3N-SRC contributions. While the
prediction for plateaus in the 2N-SRC region is robust, it is much more difficult to predict where such a plateau
should be seen for 3N-SRCs, or even if it will present itself in this fashion~\cite{Ye:2017mvo}.

\begin{figure}[htb]
\centering
\includegraphics[height=2.5in]{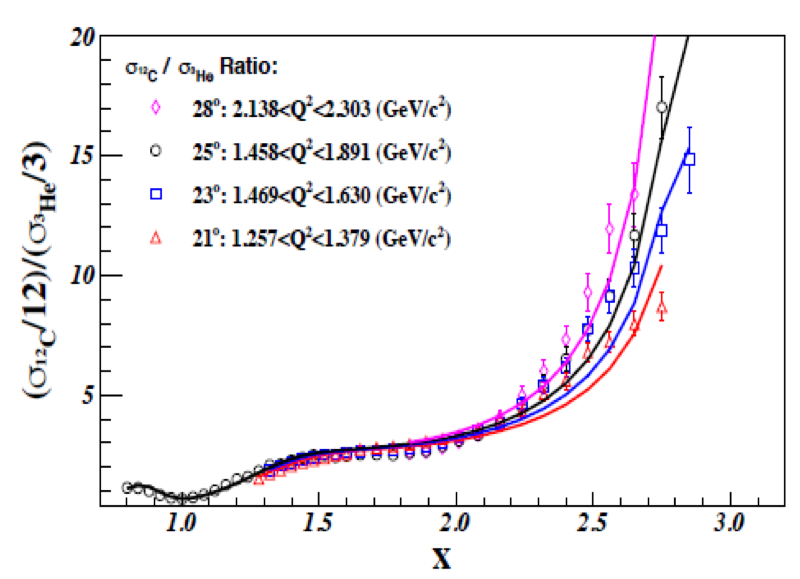}
\caption{The $^{4}$He/ $^{3}$He ratios at different $Q^{2}$ from the E08-014 data~\cite{Ye:2017mvo} }.
\label{fig:3nsrc_Q2}
\end{figure}

\begin{figure}[htb]
\centering
\includegraphics[height=2.5in]{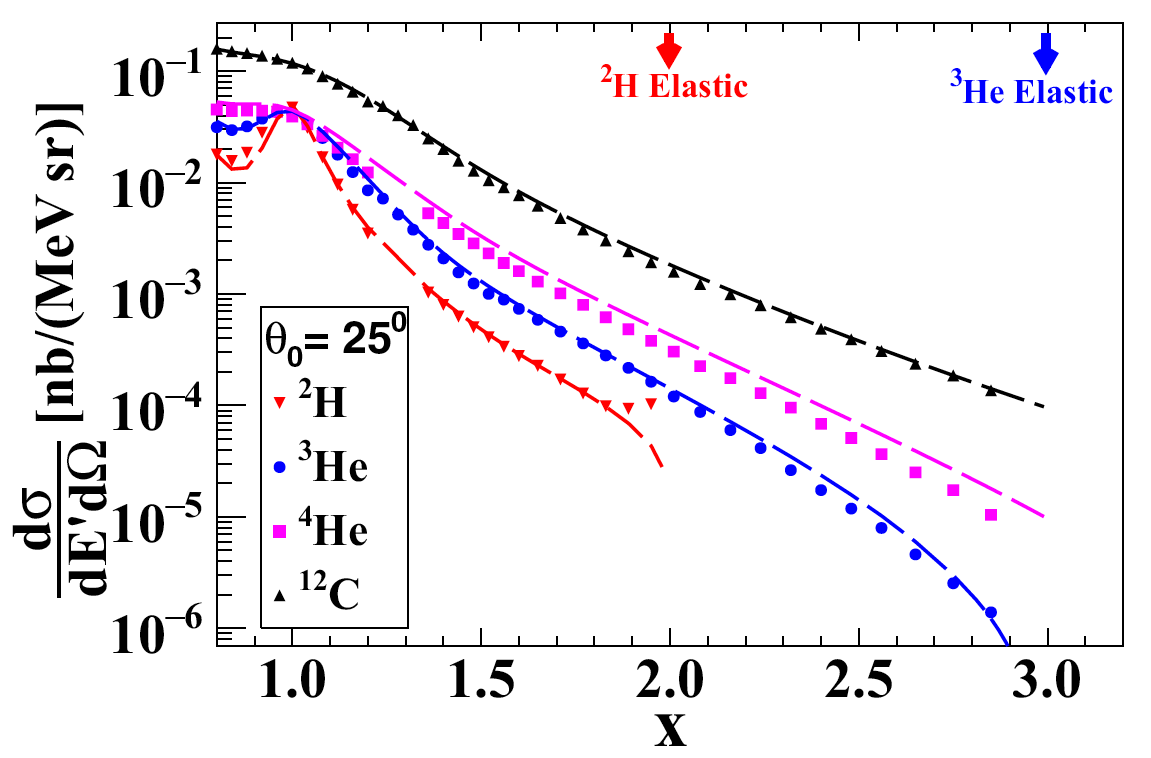}
\caption{New QE cross section results for multiple nuclei from the Hall A E08-014 experiment ~\cite{Ye:2017mvo} }.
\label{fig:xgt2_xs}
\end{figure}

This leaves us with the question of whether or not 3N-SRC are important in nuclei, and demonstrates the need for 
further theoretical and experimental work to provide a clear and quantitative statement on their contributions in 
nuclei. Isolating 2N-SRC requires $x>1.4$ and $Q^2\geq$1.5~GeV$^2$. At $x\rightarrow 2$ the scaling behavior of
the 2N-SRC
breaks down due to the strong motion of the 2N-SRC pair in a nucleus. When $x$ gets even larger, one expects that
the 2N-SRC contribution is eventually overwhelmed by the 3N-SRC contributions. However, it is not clear when this
transition happens. Some theoretical calculations, e.g.~\cite{Sargsian:2012sm}, argue that a much larger $Q^{2}$ is
needed to ($Q^{2}>10~GeV^2$) to be sensitive to the 3N-SRC for $x>2$. At the E08-014 kinematic region, the ratio
has strong $Q^{2}$ dependent as shown in Fig.~\ref{fig:3nsrc_Q2}, and the onset of the 3N-SRC may show up very
late, e.g. $x>2.5$. On the other hand, one should also note that unlike the 2N-SRC pair which always contains two
back-to-back nucleons, the momentum balance in a 3N-SRC cluster allows for a range of symmetric and asymmetric 
distributions in terms of both momentum and isospin. Fig.~\ref{fig:xgt2_xs} shows the QE cross-section distribution
from the E08-014 data
up to $x=3$ for multiple nuclei. One can clearly see that the distributions indicate scaling behavior in $1.5<x<2$
which gives the 2N-SRC plateau. When $x$ becomes larger, the $^{3}$He drops significantly faster than other nuclei
and eventually goes to zero at $x\rightarrow 3$. Compared with the deuterium cross sections, one can conclude that
the motion of the 3N-SRC has larger effect near $x=3$ which possibly cause the onset of the 3N-SRC deforms much
earlier before $x\rightarrow 3$.

\section{Summary and Future Perspective}
In the last three decades, lots of progress have been made in term of studying the nucleon-nucleon interaction in a
dramatic form where the SRC describe the feature when nucleons are largely overlapped. Many interesting results
were released from experiments at SLAC, BNL, and JLab. During the JLab 12GeV era, we have a set up of new
experiments that have been approved to continue pursue the topic of SRC using inclusive scattering and systematic
study the connection between the SRC and the EMC effect. The Hall A Tritium run group experiments, which have been
running from December 2017 to the end of 2018, will explore the isospin effect in SRC and 3N-SRC using the
inclusive QE scattering off $^{3}$He and $^{3}$H~\cite{e12-11-112}, and also study the EMC effect using deep
inelastic scattering off these two nuclei~\cite{marathon}. In Hall C, multiple experiments were carefully designed
to investigate the SRC and the EMC effects in a much broader kinematic region with a very similar set of nuclear
targets~\cite{e12-06-105, e12-10-108}.

\Acknowledgements
This work was supported by the U.S. Department of Energy, Office of Science, Office of Nuclear Physics,
under contract DC-AC02-06CH11357.

\end{document}

%% file: econfmacros.tex
\def\beq{\begin{equation}}
\def\eeq#1{\label{#1}\end{equation}}
\def\eeqn{\end{equation}}

\def\beqa{\begin{eqnarray}}
\def\eeqa#1{\label{#1}\end{eqnarray}}
\def\eeqan{\end{eqnarray}}

\let\bar=\overbar

\def\Dslash{\not{\hbox{\kern-4pt $D$}}}
\def\dslash{\not{\hbox{\kern-2pt $\del$}}}

\def\msb{{\bar{\ssstyle M \kern -1pt S}}}

